\documentclass[compsoc, conference, letterpaper, 10pt, times]{IEEEtran}
\usepackage{url}
\usepackage{stmaryrd}
\usepackage{amsmath}
\usepackage{amsfonts}
\usepackage{color}
\usepackage{amssymb}
\usepackage{graphicx}
\usepackage{tcolorbox}
\usepackage{flushend}
\begin{document}

\newcommand{\todobox}[3]{%
    \colorbox{#1}{\textcolor{white}{\sffamily\bfseries\scriptsize #2}}%
#3 %
        \textcolor{#1}{$\blacktriangleleft$}%
}


\newcommand{\todo}[1]{\todobox{red}{TODO}{#1}}
\newcommand{\feedback}[1]{\todobox{orange}{OK?}{#1}}
\newcommand{\done}[1]{\todobox{blue}{DONE}{#1}}

\renewcommand{\done}[1]{}
	
\newcommand{\bit}{\{0,1\}}
\newcommand{\bitstring}[1]{\ensuremath{\bit^{#1}}}
\newcommand{\Seq}[2]{#1;\ #2}
\newcommand{\Ass}[2]{#1 \leftarrow #2}
\newcommand{\Rand}[2]{#1 \stackrel{\raisebox{-.25ex}[.25ex]%
{\tiny $\mathdollar$}}{\raisebox{-.2ex}[.2ex]{$\leftarrow$}} #2}
\newcommand{\Randb}[1]{\Rand{#1}{\bit}}
\newcommand{\Randbs}[2]{\Rand{#1}{\bitstring{#2}}}
\newcommand{\Cond}[3]{\mathsf{if}\ #1\ \mathsf{then}\\ \hspace{1em} \ #2 \\ \mathsf{else}\\ \hspace{1em} #3\\}
\newcommand{\Condt}[2]{\mathsf{if}\ #1\ \mathsf{then}\ #2}
\newcommand{\Elsif}{\mathsf{elsif}\ }

\newcommand{\return}{\mathsf{return}\ }

\newcommand{\Prob}{\mathbb{P}}
\newcommand{\request}{\mathsf{request}}
\newcommand{\test}{\mathsf{test}}
\newcommand{\PRG}{\textsf{2PRG}}
\newcommand{\BC}{\textsf{BC}}
\newcommand{\Adv}{\textsf{Adv}}
\newcommand{\Pred}{\textsc{Pred}}
\newcommand{\Succ}{\textsc{Succ}_\Adv}
\newcommand{\Sched}{S}	

\newcommand{\secsem}[1]{\llbracket #1 \rrbracket}
\newcommand{\sem}[1]{[ #1 ]}

\newtheorem{definition}{Definition}
\newtheorem{theorem}{Theorem}
\newtheorem{example}{Example}
\newtheorem{lemma}{Lemma}
\newtheorem{corollary}{Corollary}

\title{
  Reasoning about Probabilistic Defense Mechanisms against Remote Attacks
}


\author{
    \IEEEauthorblockN{Mart\'{i}n Ochoa\IEEEauthorrefmark{1}, Sebastian Banescu\IEEEauthorrefmark{2}, Cynthia Disenfeld\IEEEauthorrefmark{3}, Gilles Barthe\IEEEauthorrefmark{4}, Vijay Ganesh\IEEEauthorrefmark{5} }
    \IEEEauthorblockA{\IEEEauthorrefmark{1}Singapore University of Technology and Design}
    \IEEEauthorblockA{\IEEEauthorrefmark{2}Technische Universit\"{a}t M\"{u}nchen}
    \IEEEauthorblockA{\IEEEauthorrefmark{3}University of Toronto}
    \IEEEauthorblockA{\IEEEauthorrefmark{4} IMDEA Software}
	 \IEEEauthorblockA{\IEEEauthorrefmark{5} University of Waterloo}	
}

\maketitle

\begin{abstract}

Despite numerous countermeasures proposed by practitioners and
researchers, remote control-flow alteration of programs with
memory-safety vulnerabilities continues to be a realistic
threat. Guaranteeing that complex software is completely free of memory-safety 
vulnerabilities is extremely expensive. Probabilistic countermeasures that 
depend on random secret keys are interesting, because they are an inexpensive 
way to raise the bar for attackers who aim to exploit memory-safety 
vulnerabilities. Moreover, some countermeasures even support legacy systems.
However, it is unclear how to quantify and compare the effectiveness of 
different probabilistic countermeasures or combinations of such countermeasures.
In this paper we propose a methodology to rigorously derive security bounds
for probabilistic countermeasures. We argue that by representing
security notions in this setting as events in probabilistic games,
similarly as done with cryptographic security definitions, concrete
and asymptotic guarantees can be obtained against realistic
attackers. These guarantees shed light on the effectiveness of single
countermeasures and their composition and allow practitioners to more
precisely gauge the risk of an attack.
\end{abstract}


\section{Introduction}

Memory-safety vulnerabilities, such as writing outside of boundaries of
buffers of programs written in C/C++, represent a high security risk for applications
facing a network or malicious input, since they can be exploited to
execute arbitrary code on a victim's machine
\cite{younan2012runtime}. Although due to its severity, several
countermeasures have been proposed in the last 15 years to mitigate
the risk of exploitation, a new generation of exploits taking
advantage of code re-use techniques
\cite{shacham2007geometry,bittau2014hacking,snow2013just} pose
challenges to practitioners and researchers. Also, as the difficulty
to exploit buffer overflows (BOF) increases, attackers resort to
exploiting other memory-safety vulnerabilities, such as format string
vulnerabilities, heap overflows and buffer over-reads (as in the case
of the recent Heartbleed exploit~\cite{durumeric2014matter}).

Security researchers have come up with many countermeasures (we
interchangeably use the terms defense mechanisms and countermeasures)
against memory-based exploitation, broadly classified as probabilistic
and non-probabilistic. Examples of non-probabilistic countermeasures
include access control protection to certain memory regions to avoid
remote code injection on the stack \cite{younan2012runtime}. This kind
of access control is currently implemented in most operating systems
and at the hardware level by most CPU vendors.  Unfortunately, such
countermeasures cannot prevent code re-use attacks such as
Return-oriented Programming (ROP)~\cite{shacham2007geometry}. In ROP
attacks, instead of injecting malicious code, attackers find sequences
of instructions in the victim's code (called \emph{gadgets}) that can
be executed in arbitrary order by injecting pointers to them on the
stack, which essentially allows running arbitrary code without violating
the access control policy on the stack. Other
non-probabilistic countermeasures, such as Control-Flow Integrity
(CFI)~\cite{abadi2005control} propose to statically compute a valid
control-flow graph of a program and enforce run-time monitors that
detect violations to this policy. Although promising, this technique
implies source-code or binary instrumentation, and can have high
performance overheads \cite{zeng2013strato}. Moreover ROP attacks that
respect the original control-flow graph can be launched even in
contexts where CFI may be enforced \cite{goktas2014out}.

Another class of countermeasures are probabilistic~
\cite{younan2012runtime}, and can intuitively be classified under the
``software diversity'' or ``moving-target'' defense mechanisms. The
core idea behind them is that there is an element of randomness (a
\emph{seed} or a \emph{key}), that changes with every execution of a
program or with every fresh boot of the host executing the program,
depending on the implementation. This nondeterminism is used with the
goal to slow-down an attacker: without any further side-channel, an
attacker would have to guess the secret in order to launch the
attack. Depending on the size of the secret, this can effectively
increase an attacker's effort to successfully carry out an attack, and
demotivate him/her. Prominently, address space layout randomization (ASLR)
\cite{bhatkar2003address} and random stack canaries
\cite{cowan1999protecting} have found their way into most modern
operating systems. Other countermeasures such as instruction-set
randomization (ISR) \cite{isetrandomization} and PointGuard
\cite{cowan2003pointguard} have
been proposed in the literature as valuable alternatives, because of
ease of implementation (easy or no binary instrumentation) and low
performance overheads. These characteristics are crucial for practical
applications, where legacy software running on resource constrained
environments such as industrial-control systems pose several
implementation challenges.

However, the effectiveness (or the lack thereof) of probabilistic 
countermeasures has been discussed in the literature for various reasons. 
First, as stated before, clever exploitation 
techniques such as ROP \cite{shacham2007geometry}, Blind ROP 
\cite{bittau2014hacking} and JITROP \cite{snow2013just} invalidate certain 
assumptions on the attacker, made by some countermeasures, rendering them less 
effective. On the other hand, side-channels attacks \cite{weiss2006known} and 
low entropy of randomization in certain architectures
\cite{shacham2004effectiveness} make them practically insecure.

In this work, we make the observation that the guarantees provided (or 
intended) by probabilistic countermeasures for memory-safety vulnerabilities 
resemble the ones of cryptographic algorithms, and as such, can be represented 
and reasoned upon using state-of-the-art cryptographic tools and techniques. In 
particular, game-based reasoning \cite{bellare2006security} has gained momentum 
in the past decade as a useful technique to formalize proof techniques 
involving probabilistic guarantees such as semantic security 
\cite{goldreich1999foundations} among others. At the core of our contribution 
is a security property definition that abstracts away from the concrete defense 
mechanism and low-level details (such as various assembly languages, processor
architectures etc.), and that can be instantiated for various countermeasures. 
In particular, our modeling allows for reasoning about composed countermeasures 
in a natural way.  We presented some of the key ideas discussed here in an
earlier paper on {\it attack-resistance of defense
  mechanisms}~\cite{DBLP:conf/ecoop/GaneshBO15}, but have considerably
improved upon and expanded these ideas in this paper.

{\bf Problem Statement.} In this work, we address the problem of 
rigorously deriving security guarantees for probabilistic countermeasures 
against remote exploitation attacks for resource constrained but otherwise 
arbitrary attackers. We are motivated to provide formal guarantees of efficacy 
for defense mechanisms because such guarantees are crucial in characterizing 
and understanding why certain defense mechanisms work as intended while others 
fail.

{\bf Contributions.} Our contributions can be summarized as follows:

\begin{itemize}

  \item We quantify the security of a probabilistic countermeasure as
    the probability of an event in a probabilistic game, similar to
    how it is done for the formal verification of cryptographic
    primitives via security games  \cite{bellare2006security,barthe2009formal}. 
	
  \item We compute bounds for popular probabilistic countermeasures
    from the literature, showing that our approach
    generalizes. Crucially, we can analyze compositions of different
    countermeasures.
    
  \item We discuss how the use of replicas, similar as in Secure 
  Multi-Execution (SME) \cite{devriese2010noninterference} and DieHard 
  \cite{diehard} can be used to close 
  certain side-channels in the implementation of probabilistic 
  countermeasures. Additionally, we show that this technique
  can conceptually make exploitation harder against many countermeasures even 
  if the keys on which the countermeasure is based are leaked.

\end{itemize}

The rest of this paper is organized as follows. Section~\ref{sec:background} 
recaps fundamentals of stack-smashing attacks, since our running examples will 
exploit BOFs, and fundamentals of game-based cryptographic proofs. 
Section~\ref{sec:approach} presents the central concepts and definitions, and 
Section~\ref{sec:case-studies} shows applications to various countermeasures. We 
show how to reason on composed countermeasures in 
Section~\ref{sec:composition}. Section~\ref{sec:side-channels} 
discusses how to plug-in leakage due to side-channels into our reasoning, and 
how this could be prevented by means of replicas. We discuss related work in 
Section~\ref{sec:rw} and conclude in Section~\ref{sec:conclusions}.

\section{Background}
\label{sec:background}
In the following, we provide some background on memory-based exploits
and probabilistic countermeasures. We focus on stack smashing attacks because 
of their popularity and criticality, and since they will serve as a basis for 
our discussion. Finally, we summarize the fundamental
concepts behind code-based cryptographic proofs.

\subsection{Stack smashing attacks}
\label{subsec:stack-smashing}
For the sake of clarity of presentation, in this work we assume that the only 
way the attacker can violate the control-flow integrity of a remotely 
accessible program is via stack smashing attacks, e.g.~code injection and code 
re-use. 
Stack smashing attacks involve writing data input by the attacker, beyond the 
declared boundaries of a statically allocated buffer, i.e.~overflowing the 
buffer. If the input of the attacker is long enough, then it will overwrite the 
control-flow information (i.e.~frame pointer and return address) of the program 
which is also stored on the stack in the vicinity of the overflowed buffer. 
Once the currently executing function of the program is finished it will 
continue execution from the instruction indicated by the overwritten return 
address. If this return address  points to a valid instruction sequence,
then the control-flow integrity of the program has been 
violated, since the program will deviate from its intended behaviour.
Otherwise, if the return address points to a sequence of bytes that 
does not represent a valid instruction according to the instruction set 
architecture (ISA) or it points to an inaccessible memory address, then the 
program crashes. 

Code injection is a type of stack smashing attack, where the attacker inserts 
the machine code s/he wants to execute in the input that is passed to the 
vulnerable program. This input is also known as the \emph{exploit payload} or 
\emph{shellcode} and it also contains other data additional to the machine code 
to be executed, e.g.~the value with which the return address is to be 
overwritten. As described in the previous paragraph, this exploit payload will 
be written to the vulnerable buffer and beyond its declared boundary. A 
crucial step of the attack is therefore to overwrite the value of the return 
address with the address where this injected code is stored on the stack.

Code re-use is similar to code injection, however, the attacker inserts a 
sequence of addresses to existing code in the process memory of the target 
program. Each of these addresses generally points to a short sequence of 
instructions ending with a return or jump instruction, which are called 
\emph{gadgets}. In between the addresses of the gadgets that the attacker 
inserts, there may also be data values which are used by the instructions of 
the gadgets (e.g.~\texttt{pop eax} takes such a data value from the stack). 
When a gadget finishes execution it will either (1) execute the 
return instruction which loads the address of the following gadget from the 
stack and continues execution, or (2) execute the jump instruction to an 
address (provided by the attacker via the exploit payload), to the following 
gadget. Again a crucial step for the attacker is to carefully craft the exploit 
payload to overwrite the return address with the address of the first gadget 
that s/he wants to execute.

\subsection{System interface}

In the previous subsection we have mentioned that the attacker
can only remotely access the system on which the vulnerable program is running.
To be more precise, we assume a synchronous communication channel such as TCP,
i.e.\ after the attacker sends an input (request), s/he receives / observes an 
output (response) from the system. The output can be either a message sent 
directly by the vulnerable program or it can be an error message sent by the 
remote system, (e.g.\ time-out), due to a crash of the vulnerable program.

\subsection{Code-based cryptographic proofs}
Modern cryptography advocates the use of rigorous security definitions
and reductionist proofs \cite{goldreich1999foundations}. In a typical
case, security definitions are captured by probabilistic experiments
in which a challenger $\mathcal{C}$ interacts with an adversary
$\mathcal{A}$, and the adversary will \emph{win} the game if his
response to the challenge computed by $\mathcal{C}$ fulfills a
well-defined winning condition. An important class of experiments are
the so-called indistinguishability games, where an adversary must
guess some bit $b$ sampled by the challenger; in this case, the
winning condition is $b=b'$, where $b'$ is the adversary's guess.
Indistinguishability games can also be expressed by means of two
experiments: in the first experiment, the challenger uses an ideal
functionality, whereas in the second experiment, the challenger uses a
real functionality. In both cases, the adversary returns a bit $b\in
\{0,1\}$. The advantage of the adversary is then defined as the
distance (i.e. absolute value of the difference) between the
probabilities that he returns 1 in the first experiment and in the
second experiments. To manage the complexity of reductionist proofs,
some cryptographers adopt the game-hopping approach, in which bounds
on the adversary's advantage are proved by means of a sequence of
transitions \cite{bellare2006security}. A further step is to write
experiments using an imperative probabilistic programming language
\cite{bellare2006security}. One can then give a formal semantics to
probabilistic programs and use this semantics as a sound basis for
rigorous mathematical reasoning~\cite{barthe2009formal}.

One central tool in the game-hopping approach is reasoning up to
failure event. In its simplest form, it considers two programs that
are syntactically identical until a boolean flag $bad$ is set (a flag
is a boolean variable that remains true once set).

\begin{definition}[Identical until bad is set] Two games $G_1$ and
  $G_2$ are identical until bad is set, if both contain a $bad$ flag
  (that is initially set to false) and they are \textit{syntactically}
  identical but for a statement $\mathsf{if~}bad=true\mathsf{~then}~S$
  in one game and $\mathsf{if~}bad=true\mathsf{~then}~T$ in the other,
  for some sequence of instructions $T$ and $S$.
\end{definition}

Equipped with the notion of \emph{identical until bad is set}, we can proceed to state the Fundamental Lemma of 
Game Playing, which is a cornerstone of the hop-based technique proposed in \cite{bellare2006security,barthe2009formal}.

\begin{theorem}[Fundamental Lemma of Game playing] Let $G_1$ and $G_2$ be two (terminating) games identical until bad is set. Let $E$ an event defined on both games and let $F$ denote the event $bad=true$. Then:

$$|\Pr{[E]}_{G_1} - \Pr{[E]}_{G_2}| \leq \Pr{[F]}_{G_i}~~~~i=1,2$$

\end{theorem}

For a proof see \cite{bellare2006security}.

\begin{example}[PRF/PRP Switching Lemma]
A simple example that illustrates the game playing principle is the Pseudo-Random Function/Pseudo-Random Permutation (PRF/PRP) switching lemma \cite{bellare2006security}. Let $Perm(n)$ be the set of all permutations over strings of $n$ bits $\{0,1\}^n$, and let $Func(n)$ the
set of all functions from $\{0,1\}^n$ to $\{0,1\}^n$. The switching lemma states that the probability of an adversary 
distinguishing sampling from the two sets above given $q$ oracle access calls to the sampling procedure is at most $\frac{q(q-1)}{2^{n+1}}$. Formally, let $A^{\pi}$ be an adversary querying an oracle that samples a random permutation $\pi \in Perm(n)$, and $A^{\rho}$ an adversary that queries an oracle sampling a random function $\rho \in Func(n)$. The 
adversary $A^P$, for $P \in \{\rho,\pi\}$ outputs a bit (0,1) after querying the random function/permutation $q$ times.
Then:

$$|\Pr[A^{\pi} = 1] - \Pr[A^{\rho} = 1]| \leq \frac{q(q-1)}{2^{n+1}}$$

We can model the two different oracles using the \emph{pWhile} language of \cite{barthe2009formal}. Here $\Rand{x}{S}$ 
is a uniform random assignment from the set $S$. We assume that the set $I$ is empty before giving oracle access 
to the adversary.

\begin{center}	\boxed{
{\small
        \begin{array}{l}
            \textbf{Proc.}~\pi \\
				\Randbs{y}{n} \\
                \Condt{y \in I}{\\ \hspace{1em} \Rand{y}{\{0,1\}^n \setminus I} }\\
				\Ass{I}{I::y}\\
            \return y
        \end{array}
}
}	\boxed{
{\small
        \begin{array}{l}
            \textbf{Proc.}~\rho \\
				\Randbs{y}{n} \\
            \return y
        \end{array}
}
}\end{center}

where $I::y$ stands for concatenation of $y$ to a list $I$.

Now, we can syntactically transform these programs into the following, semantically equivalent ones ($S_1$ has
the same output distribution as $\pi$ and $S_2$ has the same output distribution as $\rho$):

\begin{center}	\boxed{
{\small
        \begin{array}{l}
            \textbf{Proc.}~S_1 \\
				\Randbs{y}{n} \\
                \Condt{y \in I}{\\  \hspace{1em}  \Ass{bad}{true} \\ \hspace{1em} \Rand{y}{\{0,1\}^n \setminus I} }\\
				\Ass{I}{I::y}\\
            \return y
        \end{array}
}
}	\boxed{
{\small
        \begin{array}{l}
            \textbf{Proc.}~S_2 \\
				\Randbs{y}{n} \\
                \Condt{y \in I}{\\  \hspace{1em}  \Ass{bad}{true} }\\
				\Ass{I}{I::y}\\
            \return y
        \end{array}
}
}\end{center}

Since $S_1$ is identical to $S_2$ up to setting the $bad$ flag to
$true$, we can use the fundamental lemma of game playing and derive:

$$|\Pr[A^{\pi} = 1] - \Pr[A^{\rho} = 1]| \leq \frac{q(q-1)}{2^{n+1}}$$

\begin{align*}
|\Pr[A^{\pi} = 1] - \Pr[A^{\rho} = 1]| &= \\
|\Pr[A^{S_1} = 1] - \Pr[A^{S_2} = 1]| & \leq	\Pr{[bad=true]}_{S_1} \\
& \leq \frac{q(q-1)}{2^{n+1}}
\end{align*}

since the probability of $y \in I$ is the probability of a collision
in sampling $q$ random elements, which is a standard birthday paradox
bound.
\end{example}

\emph{Interpretation of bounds.}  Probabilistic experiments are always
parametrized by a security parameter $n$; for instance, the security
parameter can be related to the size of the underlying domain (in our
example, the length of the bitstrings). Thus, the advantage of the
adversary is implicitly given in terms of a function $\epsilon(n)$,
and some care is required for interpreting the results---especially
when the advantage is expressed relative to a security assumption.
One standard interpretation is asymptotic: in this setting, one
requires that adversaries are probabilistic polynomial time (PPT)
algorithms and that their advantage is negligible in the security
parameter, i.e.\, goes to zero exponentially fast. Another, and more
appropriate for our purposes, interpretation is concrete security; in
this case, one analyzes the values of $\epsilon(n)$ for chosen values
of $n$ that reflect practical scenarios.

\section{Security of probabilistic defenses as games}
\label{sec:approach}

In this section we present our formal approach on how to characterize the 
expected security guarantees for probabilistic defense mechanisms against 
memory-safety vulnerabilities and how to formalize them as code-based games. 
In the following we assume that the execution of the programs will eventually 
timeout and therefore they are terminating. Also we assume that programs 
are determinstic (by assuming their source of randomness is isolated as a 
program input). Thus, in this setting, programs can be though of as total 
functions.

\subsection{Security definition}

Let $P$ be a low-level program for an x86-like architecture (i.e. x86 
assembly). Let $\sem{P}: \{0,1\}^*  \to \{0,1\}^*$ be a semantics 
function for $P$. Let $\secsem{P}$ be 
an ideal execution that will raise an exception and crash with a distinct 
output \textsf{crash} whenever an attempt 
to exploit a memory-safety violation is detected at runtime. 
Formally there exists a set $\Omega(P) \in 2^{\{0,1\}^*}$ such that:

$$\forall \ \omega \in \Omega, \ \sem{P}(\omega) = \mathsf{crash}$$ 

Let $\mathcal{A}$ be a PPT adversary that attacks $P$, which is potentially a 
vulnerable program. 
$\mathcal{A}$ knows the program $P$
and can interact remotely with it (send inputs and can observe the respective 
outputs) by means of a machine executing it as discussed in Section~\ref{sec:background}.

\begin{definition}[Effectiveness of defense mechanism]\label{def:effectiveness}
We say that a probabilistic countermeasure $c$ is effective against remote
exploitation if the probability of attacker $\mathcal{A}$ of distinguishing
the execution of $P$ with countermeasure $c$ (denoted $\sem{P+c}$), from
the ideal execution $\secsem{P}$ is negligible on a security parameter $n$.
The adversary has oracle access to an instance of $\sem{P+c}$, which he can 
query $q$ times (where $q$ is polynomial on $n$).
Formally:

$$|\Pr[ \mathcal{A}^{\sem{P+c}} = i]| \leq \epsilon(n)$$

Where $i$ is an input to $P$ such that $i \in \Omega(P)$ and
$\sem{P+c}(i) \neq \mathsf{crash}$, for a negligible function $\epsilon$.
\end{definition}
 
Note that in this definition we abstract away from the concrete attack (i.e. 
executing a remote shell, or disabling stack execution protection), the 
vulnerability exploited (heap or stack overflow, use after free etc.) and
the countermeasure used. We rely on
the fact that a precondition for any attack is that the memory-safety violation 
goes undetected.

\subsection{Game-based modeling}

In order to instantiate Definition~\ref{def:effectiveness}, we will write 
the core of a defense countermeasure as a probabilistic program (or game), 
as well as the ideal execution $\secsem{\cdot}$.

The ideal program execution is:

\begin{center}	\boxed{
{\small
        \begin{array}{l}
            \textbf{Proc.}~\secsem{P}(i) \\
                \Cond{i \in \Omega(P)}{\Ass{o}{\mathsf{crash}}}{\ \Ass{o}{\sem{P}(i)}}
            \return o
        \end{array}
}
}\end{center}

where, for the sake of this paper, $\Omega(P)$ is the set of all inputs that will cause the program \emph{to write out of bounds of a static buffer}. In principle our abstraction can be applied for other memory safety violations 
(such as heap overflows etc.) but we leave a thorough exploration of such extensions for future work.

Clearly $$\Pr[ i \in \Omega(P) \wedge o \neq \mathsf{crash}] = 0$$ in this case.

Instead, a regular execution of a program $P$ with buffer overflow vulnerabilities (together with an empty countermeasure $\emptyset$) can be abstracted as:

\begin{center}	\boxed{
{\small
        \begin{array}{l}
            \textbf{Proc.}~\sem{P+\emptyset}(i) \\
                \mathsf{If \ } i \in \Omega(P) \ \mathsf{then} \\
                 \hspace{1em} \Ass{ra}{i.\mathsf{payload}[0]} \\ 
                \hspace{1em} \mathsf{If \ }ra \in \mathsf{Valid} \ \mathsf{then}\\
                \hspace{2em} \Ass{o}{\sem{\mathcal{M}(P,ra)}} \\
                \hspace{1em} \mathsf{else} \\
                \hspace{2em} \Ass{o}{ \mathsf{crash}} \\
                \mathsf{else}\\
                \hspace{1em} \Ass{o}{\sem{P}(i)}\\
            \return o
        \end{array}
}
}\end{center}
where $\mathsf{Valid}$ is the set of valid return addresses. $\mathcal{M}(P, 
ra)$ is the sequence of bytes in the process memory of $P$, starting from 
address $ra$. Since this input would make the ideal execution crash, it means that it writes outside of 
bounds of a memory buffer. We assume that a function $i.\mathsf{payload}[0]$ is given that extracts the part of 
the input that overwrites the return address in the stack. For simplicity we assume that this function 
always returns a value, that is, the adversary aims at overwriting the return address to perform some sort of
code injection/reuse. Certainly, in particular cases the adversary might want just to manipulate local variables 
in the proximity of the vulnerable buffer, but in this work we limit ourselves to analyzing the more 
common and usually more critical attack that involves remote code execution.

Note that if the attacker overwrites the return address with a valid address, the exact 
output in this case is determined by the 
concrete state of the memory in the victim's host machine and could potentially 
result in a successful attack (i.e. remote code execution etc.). We do not 
assume to know a probability distribution for this memory, and therefore will 
not explicitly reason on this part of the code. Instead, note that in this 
case:

{\small
$$\Pr[ i \in \Omega(P) \wedge o \neq \mathsf{crash}] = \Pr[ i \in \Omega(P) \wedge i.\mathsf{payload}[0] \in \mathsf{Valid} ] $$
}

and will depend on the knowledge that the adversary has of $\mathsf{Valid}$. 

\subsection{Proofs}

Let $c$ be a memory-safety exploitation countermeasure. In order to show its 
effectiveness, given Definition \ref{def:effectiveness}, we can leverage the 
formal semantics of games. To this end, we can proceed in several 
ways. We can reason directly in the game describing $\sem{P+c}$ and bound the 
probability $|\Pr[ \mathcal{A}^{\sem{P+c}} = i]| \leq \epsilon(n)$ as in the security 
definition. Alternatively, following the literature in cryptographic proofs, we 
can proceed to show a bound on $\epsilon(n)$ using game-based transformations, as
illustrated in the previous section with the PRF/PRP switching Lemma. 
The idea of such transformations is to start with the game describing $\sem{P+c}$ and transform 
it until reaching  $\secsem{P}$. While doing so, we can keep track of the 
consequences of our transformations ($bad$ flags) to the probability of the event we want to 
bound. Thanks to the fundamental lemma of game 
playing~\cite{bellare2006security}, we know that:

$$|\Pr{[E]}_{\sem{P+c}} - \Pr{[E]}_{\secsem{P}}| \leq \Pr{[F]}_{\secsem{P}}$$

We then leverage on the fact that by definition $\Pr[ \mathcal{A}^{\secsem{P}} 
= i] = 0$ for $i \in \Omega(P)$ and
$\sem{P+c}(i) \neq \mathsf{crash}$, and are left to bound the probability of an 
attacker in distinguishing $\sem{P+c}$ from $\secsem{P}$.

In the following section we will discuss bounds for various countermeasures 
and illustrate our technique.

\section{Preventing attacks}
\label{sec:case-studies}

In this section we show how to model
popular probabilistic countermeasures, and how to compute
the probability of an attack for countermeasures that aim at 
preventing attack by crashing attack inputs with high 
probability.

\subsection{Stack Canaries}\label{sec:canaries}

Stack canaries \cite{wagle2003stackguard} place a random value between static
variables and the return address on the stack. For efficiency, stack
canaries have typically a fixed size of one slot on the stack (4 bytes on 32-bit systems) and are not
recomputed after each function execution due to performance reasons. 
However, there is nothing prohibiting an implementation from using larger canary sizes and updating their values after each function execution.
For simplicity of illustration, we assume that the size of the canary is potentially arbitrary, and that
it is recomputed for each new function execution. In this case:

\begin{center}	\boxed{
{\small
        \begin{array}{l}
            \textbf{Proc.}~\sem{P+c}(i) \\
                \mathsf{If \ } i \in \Omega(P) \ \mathsf{then} \\
                \hspace{1em}      \Randbs{k}{n} \\
                \hspace{1em} \Ass{ca}{i.\mathsf{payload}[0]} \\ 
                 \hspace{1em} \Ass{ra}{i.\mathsf{payload}[1]} \\ 
                \hspace{1em} \mathsf{If \ } ca = k \mathsf{\ and \ } ra \in \mathsf{Valid} \ \mathsf{then}\\
                \hspace{2em} \Ass{o}{\sem{\mathcal{M}(P,ra)}} \\
                \hspace{1em} \mathsf{else} \\
                \hspace{2em} \Ass{o}{ \mathsf{crash}} \\
                \mathsf{else}\\
                \hspace{1em} \Ass{o}{\sem{P}(i)}\\
            \return o
        \end{array}
}}\end{center}


Let the event:
$$E = i \in \Omega(P) \wedge o \neq \mathsf{crash} .$$
Now, it is easy to see that:

\begin{align*}
\Pr[ E] &= \Pr[ i \in \Omega(P) \wedge i.\mathsf{payload}[0] = k\\
 & \ \qquad{} \ \wedge  i.\mathsf{payload}[1] \in \mathsf{Valid} ] 	\\
& \leq	\Pr[ i \in \Omega(P) \wedge i.\mathsf{payload}[0] = k]  \\
& \leq \Pr[i.\mathsf{payload}[0] = k] = \frac{1}{2^n}
\end{align*}

This follows because the assignment $\Randbs{k}{n}$ guarantees that $k$ is an 
independent and uniformly random variable in the space
of bitstrings of length $n$. 

\begin{figure*}[tb]
\centering
\includegraphics[width=\linewidth]{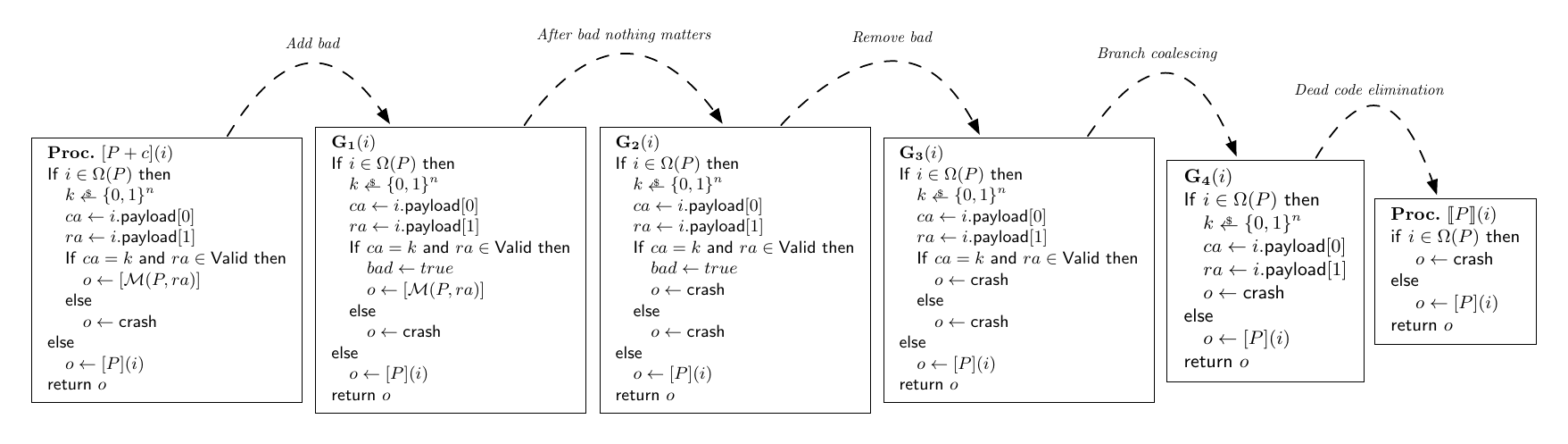}
\caption{Game transformation for a canary based countermeasure.} 
\label{fig:gamescanary}
\end{figure*}

Alternatively, we can use a sequence of game transformations as depicted 
in Fig \ref{fig:gamescanary}. We start with the game $\sem{P+c}$ which is 
semantically equivalent (same output distribution) to $\mathbf{G_1}$ that
contains a $bad$ flag in case the adversary guesses the canary and a correct 
return address. $\mathbf{G_1}$ is in turn equivalent up to $bad$ to $\mathbf{G_2}$ 
which will crash even when the adversary guesses the canary and the return 
address. After removing the bad flag ($\mathbf{G_3}$), merging the redundant branches 
($\mathbf{G_4}$) and eliminating deadcode, which are all semantics preserving 
transformations, we obtain $\secsem{P}$. 
 Applying the fundamental lemma of game playing we have:

\begin{align*}
|\Pr{[E]}_{\mathbf{G_1}} - \Pr{[E]}_{\mathbf{G_2}} | &\leq \Pr{[bad=true]}_{\mathbf{G_2}}\\
& \leq \Pr{[i.\mathsf{payload}[0] = k]_{\mathbf{G_2}}} \\
&= \frac{1}{2^n}
\end{align*}

Since in the ideal game $\Pr{[E]}_{\mathbf{G_2}} = \Pr{[E]}_{\secsem{P}} = 0$ we 
have that $\Pr{[E]}_{\sem{P+c}} = \Pr{[E]}_{\mathbf{G_1}} \leq \frac{1}{2^n} $. 
This bound is negligible in $n$ and thus we derive the security proof.

\subsubsection{Multiple sampling and single randomization}

Note that, by assumption, the adversary is a PPT. This means that he can at most query the $\sem{P+c}$ oracle a polynomial number of times $q$,
and thus the probability of finding a suitable $i \in \Omega(P)$ such that the execution is not automatically stopped is bounded by $\frac{q}{2^n}$, 
which is still a negligible function in $n$. Now, if we modify the game to better model actual implementations, and pass $k$ as a parameter to the 
execution of $P$:
\begin{center}
{\small
		\boxed{
        \begin{array}{l}
        
             \textbf{Proc.}~G_1 \\
                \hspace{1em}      \Randbs{k}{n} \\
        \end{array}} \qquad{}
		\boxed{
        \begin{array}{l}
            \textbf{Proc.}~\sem{P+c}(i) \\
                \mathsf{If \ } i \in \Omega(P) \ \mathsf{then} \\
                \hspace{1em} \Ass{ca}{i.\mathsf{payload}[0]} \\ 
                 \hspace{1em} \Ass{ra}{i.\mathsf{payload}[1]} \\ 
                \hspace{1em} \mathsf{If \ } ca = k \mathsf{\ and \ } ra \in \mathsf{Valid} \ \mathsf{then}\\
                \hspace{2em} \Ass{o}{\sem{\mathcal{M}(P,ra)}} \\
                \hspace{1em} \mathsf{else} \\
                \hspace{2em} \Ass{o}{ \mathsf{crash}} \\
                \mathsf{else}\\
                \hspace{1em} \Ass{o}{\sem{P}(i)}\\
            \return o
        \end{array}}
}\end{center}

Then the probability of finding $i \in \Omega(P) \wedge o \neq \mathsf{crash}$ after $q$ queries is given by:

$$\sum_{j=1}^{q} \frac{1}{2^n -j} \leq \frac{q}{2^n -q}$$

which is still a negligible function in $n$. Note that in practice, if the 
value of $n$ is relatively small, then the expected number of queries needed to 
find a successful guess of the canary is also small, rendering the 
countermeasure ineffective. This has been pointed 
out in \cite{shacham2004effectiveness} for ASLR, which as we will see has an 
identical bound. However, changing the key in every execution of a function has 
a huge impact in preventing attacks that rely on partial information leakage 
and \emph{side-channels}, as we will discuss in the following sections.

\subsubsection{Concrete bounds}
\label{subsec:canary-concrete-bounds}
As discussed before, in practice canaries will have often a 4 byte size
and will not be recomputed on each execution (but only on compilation). This
gives the bound\footnote{Mathematically the denominator of the concrete bounds could thus be negative for certain values of $q$. However, in practice an attacker will stop before that, because before reaching negative values the fraction will hit 1, which indicates that the attack is successful.}:

$$\sum_{j=1}^{q} \frac{1}{2^{32} -j} \leq \frac{q}{2^{32} -q}$$

This certainly is susceptible to brute-force attacks by weak adversaries. To
increase security, intuitively, one may increase the size $n$ of the canary. 
However, such an increase would have a negative impact on performance. 
If we assume the system size is $s = 32$, then $\lceil{\frac{n}{s}} \rceil$ stack slots would need to be pop-ed and compared to the canary value on each function return.
Moreover, if re-randomization is employed, a larger number of bits must be randomly generated, which also implies an increase in computation effort dependent on the type and implementation of the random number generator.

\subsection{ASLR}
\label{subsec:aslr}

ASLR \cite{shacham2004effectiveness} is a countermeasure that prevents stack exploitation by randomizing the position of the stack, heap and
code in process memory. In the following we assume that the set of valid memory addresses (position of stack, heap
and code) has constant size $|\mathsf{Valid}|$, and that the size of memory is variable and of magnitude $2^n$.
Moreover  $\Pi_k$ is a random permutation of addresses within memory. For simplicity we assume that addresses can be permuted arbitrarily: in practice there will be some constraints on the size of the program, space between heap and stack etc. Also in practice 
this randomization is done once per system start, however for simplicity we assume this is done after processing
any input:

\begin{center} \boxed{
{\small
        \begin{array}{l}
            \textbf{Proc.}~\sem{P+c}(i) \\
                \Randbs{k}{n} \\
                \mathsf{If \ } i \in \Omega(P) \ \mathsf{then} \\
                 \hspace{1em} \Ass{ra}{i.\mathsf{payload}[0]} \\ 
                \hspace{1em} \mathsf{If \ }ra \in \Pi_k(\mathsf{Valid}) \ \mathsf{then}\\
                \hspace{2em} \Ass{o}{\sem{\mathcal{M}(P,ra)}} \\
                \hspace{1em} \mathsf{else} \\
                \hspace{2em} \Ass{o}{ \mathsf{crash}} \\
                \mathsf{else}\\
                \hspace{1em} \Ass{o}{\sem{P}(i)}\\
            \return o
        \end{array}
}} \end{center}

Now, it is easy to see that:

\begin{align*}
\Pr[ i \in \Omega(P) \wedge o \neq \mathsf{crash}] & \leq \Pr[ i \in \Omega(P) \\ 
& \qquad{} \ \wedge \ i.\mathsf{payload}[0] \in \Pi_k(\mathsf{Valid})] \\			
& \leq \Pr[i.\mathsf{payload}[0] \in \Pi_k(\mathsf{Valid})] \\
& = \frac{|\mathsf{Valid}|}{2^n}	
\end{align*}

This follows because the assignment $\Randbs{k}{n}$ guarantees that $k$ is an independent an uniformly random variable in the space
of bitstrings of length $n$. 

\subsubsection{Multiple sampling and single randomization}

Similarly as in the case of canaries, $q$ queries yields a bound of $\frac{q \cdot |\mathsf{Valid}|}{2^n}$, 
which is still a negligible function in $n$. Now, if we modify the game to better model actual implementations, and pass $k$ as a parameter to the 
execution of $P$:

\begin{center} 
{\small
		\boxed{
        \begin{array}{l}
        
             \textbf{Proc.}~G_1 \\
                \hspace{1em}      \Randbs{k}{n} \\
        \end{array}
		} \qquad{}
		\boxed{
        \begin{array}{l}		
            \textbf{Proc.}~\sem{P+c}(i) \\
                \mathsf{If \ } i \in \Omega(P) \ \mathsf{then} \\
                 \hspace{1em} \Ass{ra}{i.\mathsf{payload}[0]} \\ 
                \hspace{1em} \mathsf{If \ }ra \in \Pi_k(\mathsf{Valid}) \ \mathsf{then}\\
                \hspace{2em} \Ass{o}{\sem{\mathcal{M}(P,ra)}} \\
                \hspace{1em} \mathsf{else} \\
                \hspace{2em} \Ass{o}{ \mathsf{crash}} \\
                \mathsf{else}\\
                \hspace{1em} \Ass{o}{\sem{P}(i)}\\
            \return o
        \end{array}
		}
}
\end{center}

Then the probability of finding $i \in \Omega(P) \wedge o \neq \mathsf{crash}$ after $q$ queries is given by:

$$\sum_{j=1}^{q} \frac{|\mathsf{Valid}|}{2^n -j} \leq \frac{q \cdot |\mathsf{Valid}|}{2^n -q}$$

which is still a negligible function in $n$. 

\subsubsection{Concrete bounds}

In practice, for 32-bit architectures without virtualization:

$$\sum_{j=1}^{q} \frac{|\mathsf{Valid}|}{2^{32} -j} \leq \frac{q \cdot |\mathsf{Valid}|}{2^{32} -q}$$

To increase security one has to increase the size of
the permutation space $n$. This can be achieved concretely by virtualising
memory.
\subsection{PointGuard}
PointGuard~\cite{cowan2003pointguard} is a countermeasure for BOF attacks which 
aims to overwrite pointers in any memory location of a process, i.e.~stack, 
heap and static data. PointGuard augments the C compiler by encrypting all 
pointer values that are stored in memory with a key $k$ and also adds code 
necessary to decrypt pointer values right before they are loaded into CPU 
registers (denoted $Decrypt_k$).
The only pointer values which are unencrypted during program execution, are 
those located in CPU registers. Therefore, pointers are never dereferenced 
directly from process memory, but only via registers. Since registers are not 
addressable by over-writing pointer values via BOF, the unencrypted pointer 
values are kept confidential from attackers. The key $k$ is randomly generated 
on every new process execution and it is never shared outside of the process's 
address space. If an attacker overwrites the return address on the stack with 
another address (e.g.~of shellcode or of gadgets), then this address will first 
be decrypted using $k$ and then dereferenced. This will probably lead to crash 
of the program due to accessing an invalid memory location.
To bypass PointGuard an attacker thus has to guess the value 
of $k$.

Similarly, to ASLR (see Section~\ref{subsec:aslr}), we assume that the set of 
valid memory addresses has constant size $|\mathsf{Valid}|$, and that the size 
of memory is variable and of magnitude $2^n$. Also for simplicity we first show 
the case where $k$ is sampled for every input value 
passed to the program.

\begin{center}\boxed{
{\small
	\begin{array}{l}
            \textbf{Proc.}~\sem{P+c}(i) \\
                \Randbs{k}{n} \\
                \mathsf{If \ } i \in \Omega(P) \ \mathsf{then} \\
                 \hspace{1em} \Ass{ra}{i.\mathsf{payload}[0]} \\ 
                \hspace{1em} \mathsf{If \ }Decrypt_k(ra) \in \mathsf{Valid} \ \mathsf{then}\\
                \hspace{2em} \Ass{ra_d}{Decrypt_{k}(ra)} \\
                \hspace{2em} \Ass{o}{\sem{\mathcal{M}(P,ra_d)}} \\
                \hspace{1em} \mathsf{else} \\
                \hspace{2em} \Ass{o}{ \mathsf{crash}} \\
                \mathsf{else}\\
                \hspace{1em} \Ass{o}{\sem{P}(i)}\\
            \return o
        \end{array}
}}\end{center}

$Decrypt_k$ is the decryption function (e.g.~XOR with key $k$) used by the PointGuard mechanism.
It follows that:

\begin{align*}
\Pr[ i \in \Omega(P) & \wedge o \neq \mathsf{crash}] \\
 & =  \Pr[ i \in \Omega(P) \\ 
 & \ \qquad{} \ \wedge Decrypt_k(i.\mathsf{payload}[0]) \in \mathsf{Valid} ] \\
&\leq \Pr[Decrypt_k(i.\mathsf{payload}[0]) \in \mathsf{Valid}] \\
& = \frac{|\mathsf{Valid}|}{2^n}	
\end{align*}

We can achieve this bound by assuming that an encryption under
a random key behaves as a random permutation (and therefore also its inverse).

\subsubsection{Multiple sampling and single randomization}

Similarly as before, if we modify the game to better model actual implementations, and pass $k$ as a parameter to the 
execution of $P$:

\begin{center}
{\small
		\boxed{
		\begin{array}{l}
             \textbf{Proc.}~G_1 \\
                \hspace{1em}      \Randbs{k}{n} \\
        \end{array} }\qquad{}
		\boxed{
		\begin{array}{l}
            \textbf{Proc.}~\sem{P+c}(i) \\
                \mathsf{If \ } i \in \Omega(P) \ \mathsf{then} \\
                 \hspace{1em} \Ass{ra}{i.\mathsf{payload}[0]} \\ 
                \hspace{1em} \mathsf{If \ }Decrypt_k(ra) \in \mathsf{Valid} \ \mathsf{then}\\
                \hspace{2em} \Ass{ra_d}{Decrypt_{k}(ra)} \\
                \hspace{2em} \Ass{o}{\sem{\mathcal{M}(P,ra_d)}} \\
                \hspace{1em} \mathsf{else} \\
                \hspace{2em} \Ass{o}{ \mathsf{crash}} \\
                \mathsf{else}\\
                \hspace{1em} \Ass{o}{\sem{P}(i)}\\
            \return o
        \end{array}}
}\end{center}

Then the probability of finding $i \in \Omega(P) \wedge o \neq \mathsf{crash}$ after $q$ queries is given by:

$$\sum_{j=1}^{q} \frac{|\mathsf{Valid}|}{2^n -j} \leq \frac{q \cdot |\mathsf{Valid}|}{2^n -q}$$

which is still a negligible function in $n$.

\subsubsection{Concrete bounds}

For 32-bit architectures such as x86, the size of $k$ is 32 bits. Therefore, the probability of injecting a valid address is identical to that of ASLR (see Section \ref{subsec:aslr}:

$$\sum_{j=1}^{q} \frac{|\mathsf{Valid}|}{2^{32} -j} \leq \frac{q \cdot |\mathsf{Valid}|}{2^{32} -q}$$

Increasing security in this case is not only related to the size of
the memory, but also that of the key $k$. 
This means security is not increased by increasing the size of they key beyond the size of a memory pointer.
Hence, the size of memory pointers must be increased, e.g.~by applying virtualization or switching of a 64-bit architecture.

\subsection{ISR}

Instruction Set Randomization (ISR) \cite{isetrandomization} is a countermeasure for code injection attacks via BOFs. 
Code injection attacks are different from ROP and JOP attacks, because they require the attacker to send executable shellcode as part of their input, while the ROP and JOP attacks require sending a sequence of gadget addresses and other non-executable data values.
Therefore, the code injection attacker can be considered weaker than the ROP or JOP attacker.
Nevertheless, relatively recent work \cite{papadogiannakis2013asist} has shown that ISR can be combined with stack canaries in order to defend against both attackers.
This kind of composite defense follows the \emph{defense in depth} security principle, where a ROP attacker first disables the execution prevention mechanism on code on the stack (e.g.~DEP on MS Windows, $W \oplus X$ on Linux), and then executes any code that is injected on the stack.
Employing ISR raises the bar for such attacks, therefore, we believe it is interesting to model this protection mechanism.

Similarly to PointGuard, ISR uses encryption via a key $k$ to prevent attacks. 
Differently from PointGuard it encrypts all the instructions in the code segment, instead of the pointers in all memory segments.
Whenever the CPU fetches an instruction to execute, it must be first decrypted using $k$.
Since the key $k$ must be stored in a write-only CPU register which cannot be accessed by attackers, the authors have augmented a CPU emulator with such a register.
If an attacker injects shellcode onto the stack via a BOF exploit, this shellcode will be first decrypted (denoted $Decrypt_k$, e.g.~XOR with key $k$) and then executed by the CPU.
This will highly likely lead to a crash of the program due to an undefined instruction, i.e.~the CPU will not be able to execute the decrypted instruction because it is not part of its \emph{Instruction Set Architecture} (ISA).
Therefore, an attacker can bypass ISR by guessing the correct value of $k$ and encrypting the exploit payload with it before injecting it into the vulnerable program. 
However, ISR can also be bypassed by ROP and JOP attacks~\cite{shacham2007geometry, bletsch2011jump}. 
These attacks do not inject code on the stack, instead, they inject pointers to existing code.
Since existing code is already encrypted, it will be properly executed by a program which employs ISR as a countermeasure.

\begin{center} \boxed{
{\small
		   	\begin{array}{l}
            \textbf{Proc.}~\sem{P+c}(i) \\
                \Randbs{k}{m} \\
                \mathsf{If \ } i \in \Omega(P) \ \mathsf{then} \\
                 \hspace{1em} \Ass{ra}{i.\mathsf{payload}[0]} \\ 
                \hspace{1em} \mathsf{If \ }ra \in \mathsf{Valid} \ \mathsf{then}\\
                \hspace{2em} \mathsf{If \ }Decrypt_k(\mathcal{M}(P,ra)) \in \mathsf{ISA} \ \mathsf{then}\\
                \hspace{3em} \Ass{o}{\sem{\mathcal{M}(P,ra)}} \\
                \hspace{2em} \mathsf{else} \\
                \hspace{3em} \Ass{o}{ \mathsf{crash}} \\
                \hspace{1em} \mathsf{else} \\
                \hspace{2em} \Ass{o}{ \mathsf{crash}} \\
                \mathsf{else}\\
                \hspace{1em} \Ass{o}{\sem{P}(i)}\\
            \return o
        \end{array}
}}
\end{center}

For the sake of simplicity we assume an ISA with all instructions having a fixed width of $m$-bits.
It follows that:
\begin{align*}
\Pr[ i \in \Omega(P) & \wedge o \neq \mathsf{crash}] \\
& = \Pr[ i \in \Omega(P) \\ 
& \ \quad{} \ \wedge Decrypt_k(\mathcal{M}(P,i.\mathsf{payload}[0]) \in \mathsf{ISA} ] \\
& \leq \Pr[Decrypt_k(\mathcal{M}(P,i.\mathsf{payload}[0])) \in \mathsf{ISA}] \\
& = \frac{|\mathsf{ISA}|}{2^m}	
\end{align*}

\subsubsection{Multiple sampling and single randomization}

Similarly as before, if we modify the game to better model actual implementations, and pass $k$ as a parameter to the 
execution of $P$:

\begin{center}
{\small
		\boxed{
		\begin{array}{l}
             \textbf{Proc.}~G_1 \\
                \hspace{1em}      \Randbs{k}{m} \\
        \end{array} }
}\end{center}		

\begin{center}{		
		\boxed{
		\begin{array}{l}
            \textbf{Proc.}~\sem{P+c}(i) \\
                \mathsf{If \ } i \in \Omega(P) \ \mathsf{then} \\
                 \hspace{1em} \Ass{ra}{i.\mathsf{payload}[0]} \\ 
                \hspace{1em} \mathsf{If \ }ra \in \mathsf{Valid} \ \mathsf{then}\\
                \hspace{2em} \mathsf{If \ }Decrypt_k(\mathcal{M}(P,ra)) \in \mathsf{ISA} \ \mathsf{then}\\
                \hspace{3em} \Ass{o}{\sem{\mathcal{M}(P,ra)}} \\
                \hspace{2em} \mathsf{else} \\
                \hspace{3em} \Ass{o}{ \mathsf{crash}} \\
                \hspace{1em} \mathsf{else} \\
                \hspace{2em} \Ass{o}{ \mathsf{crash}} \\
                \mathsf{else}\\
                \hspace{1em} \Ass{o}{\sem{P}(i)}\\
            \return o
        \end{array}}
}\end{center}

Then the probability of finding $i \in \Omega(P) \wedge o \neq \mathsf{crash}$ after $q$ queries is given by:

$$\sum_{j=1}^{q} \frac{|\mathsf{ISA}|}{2^m -j} \leq \frac{q \cdot |\mathsf{ISA}|}{2^m -q}$$

which is still a negligible function in $m$.

\subsubsection{Concrete bounds}

For the 32-bit fixed width ISA we assumed earlier the probability of injecting a valid instruction is:

$$\sum_{j=1}^{q} \frac{|\mathsf{ISA}|}{2^{32} -j} \leq \frac{q \cdot |\mathsf{ISA}|}{2^{32} -q}$$

Increasing security in this case is directly related to the width of the instructions in the ISA and the size of the key $k$.
However, similarly to PointGuard, increasing the size of the key beyond the width of the instructions in the ISA does not increase the security.
This can be shown in practice, where the width of instructions between and within an ISA may vary.
For instance, in the x86 architecture the smallest instructions are 8-bits, while the longest instructions are 120-bits.
This is precisely the insight that Sovarel et al.~\cite{sovarel2005s} used to guess the 32-bit key, one-byte at a time, i.e.~XOR-ing an 8-bit instruction with a 32-bit key, only uses 8-bits of the key.
The search space for guessing a valid 8-bit instruction in the x86 ISA is reduced from $2^{32}$ to $2^8$.
After guessing these 8 bits of the key, the attacker can use a 16-bit instruction to guess the next 8 bits, a.s.o.~until the entire 32-bits of the key are guessed.
Note that this attack only works in the case of multiple sampling and single randomization, i.e.~if the key is changed on every input, then the attack has a much lower success rate.
The attack by Sovarel et al.~\cite{sovarel2005s} is a prime example of what can go wrong with randomization based defenses if the key size is picked independently of the range of the data items that it must protect.

\section{Composition}
\label{sec:composition}

In practice, in order to raise the bar against attackers, various defense mechanisms are usually stacked together.
In the following we discuss examples of composed systems, how to modeal and reason about them using 
games and their resulting bounds.

\subsection{ASLR $\otimes$ Canaries}
ASLR and Canaries are commonly used simultaneously in modern systems.
We model the composed countermeasure $c=$ ASLR $\otimes$ Canaries as follows:

\begin{center}\boxed{
{\small
        \begin{array}{l}
            \textbf{Proc.}~\sem{P+c}(i) \\
                \mathsf{If \ } i \in \Omega(P) \ \mathsf{then} \\
                \hspace{1em}      \Randbs{k_1}{n} \\
                \hspace{1em}      \Randbs{k_2}{m} \\
                \hspace{1em} \Ass{ca}{i.\mathsf{payload}[0]} \\ 
                 \hspace{1em} \Ass{ra}{i.\mathsf{payload}[1]} \\ 
                \hspace{1em} \mathsf{If \ } ca = k_1\ \mathsf{then}\\
                \hspace{2em} \mathsf{If \ } ra \in \Pi_{k_2}(\mathsf{Valid}) \ \mathsf{then}\\
                \hspace{3em} \Ass{o}{\sem{\mathcal{M}(P,ra)}} \\
                \hspace{2em} \mathsf{else} \\
                \hspace{3em} \Ass{o}{ \mathsf{crash}} \\
                \hspace{1em} \mathsf{else} \\
                \hspace{2em} \Ass{o}{ \mathsf{crash}} \\
                \mathsf{else}\\
                \hspace{1em} \Ass{o}{\sem{P}(i)}\\
            \return o
        \end{array}
}}\end{center}

This modeling is motivated by the fact that when composed, an attacker must bypass both the
canary and the ASLR protection in order to avoid a crash.

Now, we can calculate the following bound:

\begin{align*}
\Pr[ i \in \Omega(P) \wedge o \neq \mathsf{crash}] & \leq \Pr[ i.\mathsf{payload}[0] = k_1 \\
& \ \qquad{} \wedge \ i.\mathsf{payload}[1] \in \Pi_{k_2}(\mathsf{Valid})] \\ 
& \leq \frac{1}{2^n} \cdot \frac{|\mathsf{Valid}|}{2^m}	
\end{align*}

This bound indicates that in some cases by combining probabilistic defense mechanisms, we can increase the bounds of attack resistance.
However, not all possible combinations will provide an increase.
For instance, combining ASLR and PointGuard will not increase the bounds given by PointGuard as we will
discuss later in more detail.

\subsubsection{Multiple sampling and single randomization}

Similarly as in previous examples, if we modify the game to better model actual implementations, and pass $k_1$ and $k_2$ as parameters to the 
execution of $P$:

\begin{center}
{\small
		\boxed{
		\begin{array}{l}
             \textbf{Proc.}~G_1 \\
                \hspace{1em}      \Randbs{k_1}{n} \\
                \hspace{1em}      \Randbs{k_2}{m} \\
        \end{array} }\qquad{}
		\boxed{
		\begin{array}{l}
            \textbf{Proc.}~\sem{P+c}(i) \\
                \mathsf{If \ } i \in \Omega(P) \ \mathsf{then} \\
                \hspace{1em} \Ass{ca}{i.\mathsf{payload}[0]} \\ 
                 \hspace{1em} \Ass{ra}{i.\mathsf{payload}[1]} \\ 
                \hspace{1em} \mathsf{If \ } ca = k_1\ \mathsf{then}\\
                \hspace{2em} \mathsf{If \ } ra \in \Pi_{k_2}(\mathsf{Valid}) \ \mathsf{then}\\
                \hspace{3em} \Ass{o}{\sem{\mathcal{M}(P,ra)}} \\
                \hspace{2em} \mathsf{else} \\
                \hspace{3em} \Ass{o}{ \mathsf{crash}} \\
                \hspace{1em} \mathsf{else} \\
                \hspace{2em} \Ass{o}{ \mathsf{crash}} \\
                \mathsf{else}\\
                \hspace{1em} \Ass{o}{\sem{P}(i)}\\
            \return o
        \end{array}}
}\end{center}

Then the probability of finding $i \in \Omega(P) \wedge o \neq \mathsf{crash}$ after $q$ queries trying to guess $k_1$ and $r$ queries trying to guess $k_2$ is given by:

$$\left(\sum_{j=1}^{q} \frac{1}{2^n -j}\right) \cdot \left(\sum_{l=1}^{r} \frac{|\mathsf{Valid}|}{2^m -j} \right) \leq \frac{q}{2^n -q} \cdot \frac{r \cdot |\mathsf{Valid}|}{2^m -r})$$

which is still a negligible function in $n, m$.

\subsubsection{Concrete bounds}

In practice, for 32-bit architectures without virtualization:

$$\left( \sum_{j=1}^{q} \frac{1}{2^{32} -j} \right) \cdot \left( \sum_{l=1}^{r} \frac{|\mathsf{Valid}|}{2^{32} -l} \right) \leq \frac{q}{2^{32} -q} \cdot \frac{r \cdot |\mathsf{Valid}|}{2^{32} -r}$$

\subsection{PointGuard $\otimes$ ISR}

We model the composition $c=$ PointGuard $\otimes$ ISR as follows:

\begin{center}\boxed{
{\small
        \begin{array}{l}
            \textbf{Proc.}~\sem{P+c}(i) \\
                \mathsf{If \ } i \in \Omega(P) \ \mathsf{then} \\
                \hspace{1em}      \Randbs{k_1}{n} \\
                \hspace{1em}      \Randbs{k_2}{m} \\
                \hspace{1em} \Ass{ra}{i.\mathsf{payload}[0]} \\ 
                \hspace{1em} \mathsf{If \ }Decrypt_{k_1}(ra) \in \mathsf{Valid} \ \mathsf{then}\\
                \hspace{2em} \Ass{ra_d}{Decrypt_{k_1}(ra)} \\
                \hspace{2em} \mathsf{If \ }Decrypt_{k_2}(\mathcal{M}(P,ra_d)) \in \mathsf{ISA} \ \mathsf{then}\\
                \hspace{3em} \Ass{o}{\sem{\mathcal{M}(P,ra_d)}} \\
                \hspace{2em} \mathsf{else} \\
                \hspace{3em} \Ass{o}{ \mathsf{crash}} \\
                \hspace{1em} \mathsf{else} \\
                \hspace{2em} \Ass{o}{ \mathsf{crash}} \\
                \mathsf{else}\\
                \hspace{1em} \Ass{o}{\sem{P}(i)}\\
            \return o
        \end{array}
}}\end{center}

This composition has the advantage that it is attack resistant against both code injection attacks, ROP and JOP attacks, because ISR defends against executing injected code and PointGuard defends against overwriting the return address.
We can also perform a similarly attack resistant composition between ISR and ASLR, or ISR and Canaries, or even composing more than two countermeasures.
However, we do not show these compositions here due to space restrictions.
We can calculate the following bound for PointGuard $\otimes$ ISR:

\begin{align*}
\Pr & [ i \in \Omega(P) \wedge o \neq \mathsf{crash}] \\
&\leq \Pr[ Decrypt_{k_1}(i.\mathsf{payload}[0]) \in \mathsf{Valid} \\
& \ \wedge \ Decrypt_{k_2}(\mathcal{M}(P, Decrypt_{k_1}(i.\mathsf{payload}[0])) \in \mathsf{ISA}] \\ 
&\leq \frac{|\mathsf{Valid}|}{2^n} \cdot \frac{|\mathsf{ISA}|}{2^m}	
\end{align*}

Note that the attacker has to not only enter a validly encrypted return address with key $k_1$ but also to inject a validly encrypted sequence of instructions at that address.

\subsubsection{Multiple sampling and single randomization}

If we consider implementations of PointGuard and ISR where new keys are generated for each program input, then the model becomes:

\begin{center}
{\small
		\boxed{
		\begin{array}{l}
             \textbf{Proc.}~G_1 \\
                \hspace{1em}      \Randbs{k_1}{n} \\
                \hspace{1em}      \Randbs{k_2}{m} \\
        \end{array} }\qquad{}}
\end{center}

\begin{center}	
	{\small			
		\boxed{
		\begin{array}{l}
            \textbf{Proc.}~\sem{P+c}(i) \\
                \mathsf{If \ } i \in \Omega(P) \ \mathsf{then} \\
                \hspace{1em} \Ass{ra}{i.\mathsf{payload}[0]} \\ 
                \hspace{1em} \mathsf{If \ }Decrypt_{k_1}(ra) \in \mathsf{Valid} \ \mathsf{then}\\
                \hspace{2em} \Ass{ra_d}{Decrypt_{k_1}(ra)} \\
                \hspace{2em} \mathsf{If \ }Decrypt_{k_2}(\mathcal{M}(P,ra_d)) \in \mathsf{ISA} \ \mathsf{then}\\
                \hspace{3em} \Ass{o}{\sem{\mathcal{M}(P,ra_d)}} \\
                \hspace{2em} \mathsf{else} \\
                \hspace{3em} \Ass{o}{ \mathsf{crash}} \\
                \hspace{1em} \mathsf{else} \\
                \hspace{2em} \Ass{o}{ \mathsf{crash}} \\
                \mathsf{else}\\
                \hspace{1em} \Ass{o}{\sem{P}(i)}\\
            \return o
        \end{array}}
}\end{center}

In this case, the probability of finding an input that violates memory safety $i \in \Omega(P)$, after $q$ queries trying to guess $k_1$ and $r$ queries trying to guess $k_2$, is equal to:
$$\left(\sum_{j=1}^{q} \frac{|\mathsf{Valid}|}{2^n -j}\right) \cdot \left(\sum_{l=1}^{r} \frac{|\mathsf{ISA}|}{2^m -j} \right) \leq \frac{q \cdot |\mathsf{Valid}|}{2^n -q} \cdot \frac{r \cdot |\mathsf{ISA}|}{2^m -r},$$
which is a negligible function in $n, m$.

\subsubsection{Concrete bounds}

For 32-bit architectures where the ISA has a fixed width, the bounds are equal to:
$$\left(\sum_{j=1}^{q} \frac{|\mathsf{Valid}|}{2^{32} -j}\right) \cdot \left(\sum_{l=1}^{r} \frac{|\mathsf{ISA}|}{2^{32} -j} \right) \leq \frac{q \cdot |\mathsf{Valid}|}{2^{32} -q} \cdot \frac{r \cdot |\mathsf{ISA}|}{2^{32} -r},$$

\begin{table*}
 \centering
 \begin{tabular}{|l|l|r|r|r|}
  \hline
  \textbf{Composition} & \textbf{$n$-bit Architecture} & \textbf{32-bit Architecture} & \textbf{64-bit Architecture} & \textbf{128-bit Architecture}\\ \hline
  \hline
  ASLR$\otimes$PointGuard & $\frac{q\cdot |\mathsf{Valid}|}{2^{n}-q}$ & 1 & $2^{-23}$ & $2^{-87}$ \\ \hline
  ASLR$\otimes$ISR & $\frac{q\cdot |\mathsf{Valid}|}{2^{n}-q}\cdot\frac{r\cdot|\mathsf{ISA}|}{2^{n}-r}$ & $2^{-10}$ & $2^{-75}$ & $2^{-203}$ \\ \hline
  PointGuard$\otimes$ISR & $\frac{q\cdot |\mathsf{Valid}|}{2^{n}-q}\cdot\frac{r\cdot|\mathsf{ISA}|}{2^{n}-r}$ & $2^{-10}$ & $2^{-75}$ & $2^{-203}$ \\ \hline
  Canary$\otimes$ASLR & $\frac{q}{2^{n}-q}\cdot\frac{r\cdot|\mathsf{Valid}|}{2^{n}-r}$ & $2^{-22}$ & $2^{-87}$ & $2^{-215}$ \\ \hline
  Canary$\otimes$PointGuard & $\frac{q}{2^{n}-q}\cdot\frac{r\cdot|\mathsf{Valid}|}{2^{n}-r}$ & $2^{-22}$ & $2^{-87}$ & $2^{-215}$ \\ \hline
  Canary$\otimes$ISR & $\frac{q}{2^{n}-q}\cdot\frac{r\cdot|\mathsf{ISA}|}{2^{n}-r}$ & $2^{-26}$ & $2^{-91}$ & $2^{-219}$ \\ \hline \hline
  ASLR$\otimes$PointGuard$\otimes$ISR & $\frac{q\cdot |\mathsf{Valid}|}{2^{n}-q}\cdot\frac{r\cdot|\mathsf{ISA}|}{2^{n}-r}$ & $2^{-10}$ & $2^{-75}$ & $2^{-203}$\\ \hline
  Canary$\otimes$ASLR$\otimes$PointGuard & $\frac{q}{2^{n}-q}\cdot\frac{r\cdot|\mathsf{Valid}|}{2^{n}-r}$ & $2^{-22}$ & $2^{-87}$ & $2^{-215}$ \\ \hline
  Canary$\otimes$ASLR$\otimes$ISR & $\frac{q}{2^{n}-q}\cdot\frac{r\cdot|\mathsf{Valid}|}{2^{n}-r} \cdot \frac{t \cdot |\mathsf{ISA}|}{2^{n}-t}$ & $2^{-42}$ & $2^{-139}$ & $2^{-331}$ \\ \hline
  Canary$\otimes$PointGuard$\otimes$ISR & $\frac{q}{2^{n}-q}\cdot\frac{r\cdot|\mathsf{Valid}|}{2^{n}-r} \cdot \frac{t \cdot |\mathsf{ISA}|}{2^{n}-t}$ & $2^{-42}$ & $2^{-139}$ & $2^{-331}$ \\ \hline
  Canary$\otimes$ASLR$\otimes$PointGuard$\otimes$ISR & $\frac{q}{2^{n}-q}\cdot\frac{r\cdot|\mathsf{Valid}|}{2^{n}-r} \cdot \frac{t \cdot |\mathsf{ISA}|}{2^{n}-t}$ & $2^{-42}$ & $2^{-139}$ & $2^{-331}$ \\ \hline
 \end{tabular}
 \vspace{1em}
 \caption{Concrete bounds for countermeasure compositions}
 \label{tab:composition-overview}
\end{table*}

\subsection{Overview of compositions}

According to AT\&T\footnote{\url{http://ipnetwork.bgtmo.ip.att.net/pws/network_delay.html}} the current overall average network latency is 33 ms and no inter-USA state latency is faster than 4 ms.
Therefore, in the worst case we assume that a remote attacker to be able to submit attack requests (queries) with a speed of 1 query/ms.
Given this optimistic attack speed of 1 query/ms we also assume that after a total number of $2^{25}$ queries, a network administrator or an intrusion prevention system will block the attacker.
This translates into over 9 days of continuous queries sent by the attacker, which again aims to depict a worst case scenario.

We do not explicitly model the remaining possible compositions of countermeasures presented in Section~\ref{sec:case-studies}.
However, in Table~\ref{tab:composition-overview} we provide an overview of both the general formulas and the concrete bounds of the remaining compositions, for 32-bit, 64-bit and virtualized 128-bit architectures, where $q$, $r$ and $t$ represent the number of guesses for the different keys of the composed countermeasures.
Note that in order to obtain the values in this table we also considered the fact that the size of the ISA is $2^{12}$ in the worst case, because even in the Intel x86 ISA we have 3683 instructions, if we consider all possible mnemonics and operand types \cite{x86_instr_count}.
Finally, we assume the size of $|\mathsf{Valid}|$ is $2^{16}$, because according to Follner et al.~\cite{follner2016analyzing}, the number of ROP gadgets in a programs range from a few hundred in programs as small as bzip2, to a little under 60.000 ROP gadgets for programs such as GCC.
For many practical purposes the number of ROP gadgets can be a useful estimation of $|\mathsf{Valid}|$, however, if one wants to use a sound upper bound the total number of instructions can be counted.
Since our paper assumes that ISAs have a fixed width equal to the size of one word, the size of $|\mathsf{Valid}| = 2^{16}$  corresponds roughly to $2^{18}$, $2^{19}$ and $2^{20}$ byte programs (ca.~256 KBs, 512 KBs and 1 MBs) for 32-, 64- and 128-bit architectures, respectively.

Since these countermeasures target different parts of the program (e.g.~instructions for ISR, pointers for PointGuard and stack memory for Canaries), or are applied at different points of the program lifetime (e.g.~compile-time for ISR, PointGuard and Canaries or load-time for ASLR), the order in which these countermeasures is fixed, i.e.~the order of applying the countermeasures cannot be changed.

Note that any composition involving PointGuard and ASLR takes the same bounds of either one of the countermeasures.
This is due to the fact that both countermeasures force attackers to guess a valid memory address.
On the other hand, ISR and Canaries force attackers to guess the key used to encrypt instructions and the canary value, respectively.
Therefore, the lowest bound obtainable from any composition involves ISR, Canaries and either PointGuard or ASLR.

\section{Side-channel attacks}
\label{sec:side-channels}

An attacker can use side-channel attacks to get information about a secret key by exploiting physical observations such as time, power consumption, noise, etc. 
For instance, if re-randomization is not employed, canaries are susceptible to brute force attacks using crashing 
as a side-channel~\cite{pi32010scraps}.
The key insight employed by this attack is that any integer number of bytes can be written passed the bounds of a buffer, (not necessarily a multiple of the machine word size). So far, our modeling assumes that the 
adversary overwrites both the canary and the return address entirely. However, in practice s/he may choose to only 
partially overwrite the canary, to learn it byte by byte, which is computationally much faster than guessing 
all bytes at once. The attacker uses the fact that the program crashes, as an oracle for his guesses.

In other words, this means that the attacker can overwrite the vulnerable buffer, up to and including only the first 
byte of the canary. If this overwritten byte is incorrect, then the program crashes, otherwise, the first byte of the canary was guessed correctly and the attacker can move on to guessing the second byte of the canary.
This side-channel attack reduces the search space from $2^k$ to $2^8 \times \lceil{\frac{n}{8}} \rceil$, which is a linear function of the key size $n$.
To be able to capture this attack, we model canaries more accurately in the following way:

\begin{center}
{\small
		\boxed{
        \begin{array}{l}
        
             \textbf{Proc.}~G_1 \\
                \hspace{1em}      \Randbs{k}{n} \\
        \end{array}} 
}\end{center}

\begin{center}{
		\boxed{
        \begin{array}{l}
            \textbf{Proc.}~\sem{P+c}(i) \\
                \mathsf{If \ } i \in \Omega(P) \ \mathsf{then} \\
                \hspace{1em} \Ass{ca}{i.\mathsf{payload}[0]} \\ 
                 \hspace{1em} \Ass{ra}{i.\mathsf{payload}[1]} \\ 
                \hspace{1em} \mathsf{If \ } ow(k,ca) = k \mathsf{\ and \ } ra \in \mathsf{Valid} \ \mathsf{then}\\
                \hspace{2em} \Ass{o}{\sem{\mathcal{M}(P,ra)}} \\
                \hspace{1em} \mathsf{else} \\
                \hspace{2em} \Ass{o}{ \mathsf{crash}} \\
                \mathsf{else}\\
                \hspace{1em} \Ass{o}{\sem{P}(i)}\\
            \return o
        \end{array}}
}\end{center}
where $ow(a,b)$ returns a string of length $|ow(a,b)|=|a|$ and overwrites the first $|b|$ bytes of $a$ with the corresponding values in $b$, leaving the remaining values identical to the corresponding positions in $a$:

\[
    ow(a,b)[i]= 
\begin{cases}
    b[i] \text{ if } 0 < i \leq |b|\\
    a[i] \text{ otherwise}
\end{cases}
\]

Therefore, if the the first entry of the payload ($i.\mathsf{payload}[0]$) is 1 byte in size, then $ow(k,ca)$ will only overwrite the first byte of the canary $k$.

\subsection{Reasoning about side-channel leakage}

We can formally plug the leakage $\lambda$ learned by an adversary to our bounds by decreasing the entropy on the key 
(originally $n$ bits) by $\lambda$ bits. For instance, if there is a side-channel leaking $\lambda$ bits of 
information on the ASLR key to the adversary after $q_1$ observations, and then an attacker performs $q_2$ subsequent observations, then our bound becomes:

\begin{align*}
\Pr[ i \in \Omega(P) \wedge o \neq \mathsf{crash}] & \leq \frac{q_2 \cdot |\mathsf{Valid}|}{2^{n-\lambda} -q_2}	
\end{align*}

Depending on how $\lambda$ increases with respect to $q$ and $n$, this could have practical relevance for concrete 
scenarios. 

\subsubsection{Concrete bounds}
Take the side-channel attack on stack canaries presented at the beginning of Section~\ref{sec:side-channels}.
If we consider $q_1 = 256 = 2^8$, i.e.\ 256 guesses made by the attacker on the first byte of the canary, then we can 
consider $\lambda = 8$-bits, since the attacker has guessed the first byte of the canary, reducing the search space by 
that amount. Note that in practice he needs in average just 128 tries to guess the first byte, here we consider the 
worst case.
Since the size of the canary can be assumed to be 4 bytes (i.e.\ $n = 32$) as we discussed in Section~\ref{subsec:canary-concrete-bounds}, the bound on a successful attack becomes:

\begin{align*}
\Pr[ i \in \Omega(P) \wedge o \neq \mathsf{crash}] & \leq \frac{1}{2^{24}}\\
\end{align*}
which is at least 4 orders of magnitude higher than $\frac{1}{2^{32}}$. If afterwards the attacker performs $q_2$ queries, this bounds becomes:

\begin{align*}
\Pr[ i \in \Omega(P) \wedge o \neq \mathsf{crash}] & \leq \frac{q_2}{2^{24} - q_2}\\
\end{align*}


In general, it is difficult to foresee and prevent all possible side-channels, in particular if they 
consider time or power consumption. However we can prevent many side-channels that are reflected directly in the values 
sent to the attacker (such as crashes and termination, or memory leaks such as Heartbleed) by relying on simultaneous 
executions of the victim program, as presented in Section~\ref{subsec:sme}.

\subsection{Probabilistic countermeasures and replicas}
\label{subsec:sme}

The use of replicas (similarly to Secure Multi-Execution, 
SME~\cite{devriese2010noninterference})  can prevent such information leakages. 
For instance, in the case of canaries, if there are at least two replicas with 
a different canary each, the probability of an attacker overriding both with 
the same values is the probability of both having the same bytes:
If the canary consists of 1 byte, then the probability of that byte 
(pseudo-randomly generated) being the same in both replicas $r_0$ and $r_1$ is 
$\Pr  [cb^{r_0} = cb^{r_1}] = \frac{1}{2^8}$. If the canary consists of $n$ 
bytes, then 
 $\Pr  [\bigwedge_{i=0}^{n-1} cb_i^{r_0} = cb_i^{r_1}] = \frac{1}{2^{8n}}$.

This reflects precisely the idea behind SME, instead of having a direct mapping for each input of the system whether it has high or low-level security, the security level is given by the nature of the input given. Valid outputs are considered of a high security level and the “crash” output due to inconsistent replicas is considered as having a low security level. Then, for a valid input (high security level), all its valid outputs are calculated and given to the user (since they also have a high security level). However, if it’s invalid (low security level), with high probability the program will crash (which will be the result of the low-level security execution.

\begin{figure}[tb]
\centering
\includegraphics[width=\linewidth]{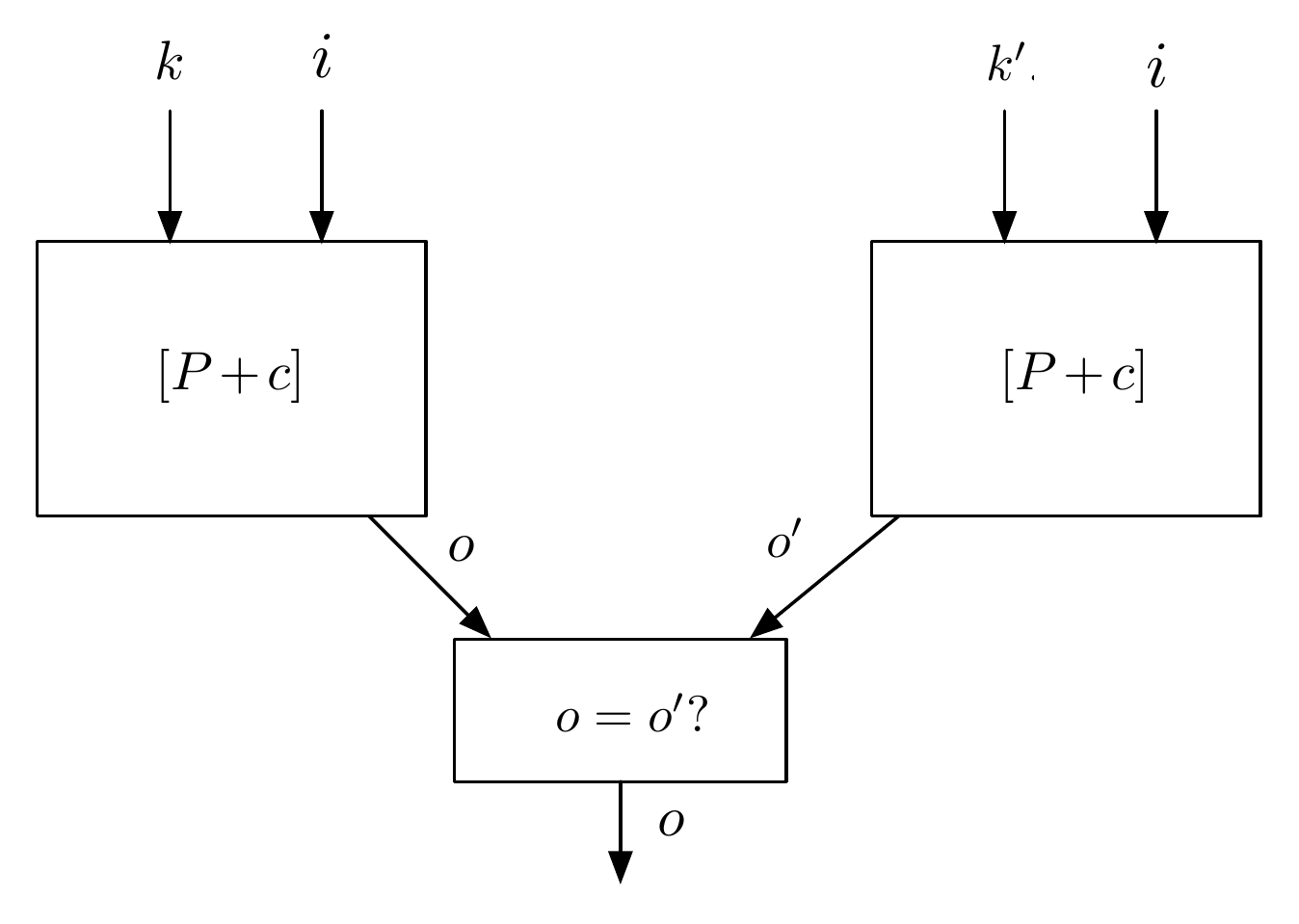}
\caption{SME inspired replicas to close value-based side-channels on $\sem{P+c}$.} 
\label{fig:sme}
\end{figure}

We can generalize this idea as depicted in Fig. \ref{fig:sme}. For any 
countermeasure relying on a secret key,
we execute two copies of the program for different keys $k$ and $k'$. Before 
sending the resulting output to the  
remote adversary, we compare the resulting outputs $o=\sem{P+c}_k[i]$ and 
$o'=\sem{P+c}_{k'}[i]$. We assume that both programs use otherwise the same 
seeds for randomness and if they interact with the system they read the same 
values (such as the system's clock, values of files, sockets etc.). If they are 
different, it means that the output is 
dependent on the only different input for the two copies: the secret keys. In 
this case, we send a \textsf{crash} 
message to the adversary and reset the program's copies with two fresh keys. 
However, \emph{legal} inputs, that is
$j \not\in \Omega$ will be in general independent from random keys used by the 
probabilistic countermeasure, thus 
this check is transparent for such inputs.

\subsection{Hardening probabilistic countermeasures with replicas} 

We note that, not only it is possible to effectively close many side-channels 
using SME, but we can also make the probabilistic countermeasures even more 
effective. In particular, we make exploitation attacks impossible in the 
following way.

\begin{lemma}
Let $\sem{P+c}$ be a probabilistic countermeasure against remote attacks. Let a 
prerequisite for exploitation that the attacker produces a payload that matches 
a random key. Formally, $i.\mathsf{payload}[0] = k$ for a deterministic oracle 
\textsf{payload}. Then, for a concrete instance of such a countermeasure using 
key $k$, an 
input $i \in \Omega(P)$ such that $\sem{P+c}_k[i] \neq \mathsf{crash}$ must 
crash another instance of the countermeasure with a key $k' \neq k$: 
$\sem{P+c}_{k'}[i] = \mathsf{crash}$.
\end{lemma}

The proof follows simply by the fact that $k \neq k'$ and that the $\mathsf{payload}$ oracle is deterministic. From this lemma, it follows:

\begin{corollary}
Let $\mathsf{SME}(\sem{P+c})$ be an implementation of $\sem{P+c}$ as described in Section~\ref{subsec:sme} and in Fig. \ref{fig:sme}, that is,
it executes two instances of $\sem{P+c}$ in parallel with random keys $k,k'$ such that $k\neq k'$, and  
outputs $\mathsf{crash}$ whenever $\sem{P+c}_k \neq \sem{P+c}_{k'}$ . Then:
$$\Pr[\mathcal{A}^{\mathsf{SME}(\sem{P+c})}=i] = 0$$
where $i\in \Omega(P)$ and $\mathsf{SME}(\sem{P+c})(i) \neq \mathsf{crash}$.
\end{corollary}

The proof of the corollary follows immediately from the lemma above. Surprisingly, the keys $k$ and $k'$ 
can be fully leaked to the attacker in this setting, since (mathematically) it 
is impossible to produce an input that would 
bypass both instances of the countermeasure.

Although this is a strong result abstractly speaking, from the point of view 
of practical implementations it suffers from several drawbacks. First, it is questionable whether executing 
a program twice will be faster than embedding a deterministic defense such as a run-time monitor. On the other hand,
depending on how the parallel execution is implemented, it could be that an attacker could craft an attack on one 
instance that disables the check on the second instance. In principle it is possible to avoid such attacks for instance 
using virtualization.

One can argue however that this idea is interesting in settings where for 
legacy reasons it is not possible to 
recompile the code with memory safety checks and where there are no hard time 
constraints on the execution, but a 
thorough efficiency and efficacy analysis is left to future work.

\section{Related work}
\label{sec:rw}

The question on the effectiveness of probabilistic countermeasures against 
memory-safety vulnerabilities has been a matter of discussion in the literature 
for the past decade. In~\cite{shacham2004effectiveness}, Schacham et al. 
discuss the effectiveness of Address Space-Layout Randomization. They noted 
that in 32 bits architectures, the entropy achieved by ASLR only slightly slows 
down attackers, while causing performance overheads. They also discuss the 
increase in security by re-randomization, but different from us, they do this 
informally and only for ASLR.

Pucella et al.~\cite{pucella2010independence} propose to treat the 
effectiveness of program diversification (i.e. using probabilistic 
transformations) as a probabilistic dynamic type-checking problem. They 
illustrate their approach on a C-like language and ASLR. Their approach is 
formal but involves the modeling of low-level details of programs, such as 
pointers, whereas our approach abstracts away from concrete programs and focus 
on countermeasures. Also, they do not explicitly give bounds on the probability 
of an attack for a given program transformation.

Abadi et al.~\cite{abadi2012protection} cast the problem of reasoning on 
the effectiveness of ASLR as full-abstraction problem. Different from Pucella 
et al., they consider concrete probability bounds on attackers, and abstract 
away from malicious inputs by considering arbitrary execution contexts. To this 
end, they also construct exemplary high and low level languages, and limit 
themselves to the analysis of ASLR.

Berger et al.~\cite{diehard} propose to randomize the 
heap layout and allocation strategy for increased resilience against memory 
management bugs that are triggered accidentally. This line of work has been 
extended~\cite{novark2010dieharder} by considering 
attackers that deliberately exploit vulnerabilities (using for instance 
heapspray attacks). Probabilistic security bounds are then derived by 
reasoning on the proposed countermeasures and successful attack events.
Such bounds, although mathematically justified, are obtained informally 
(i.e. without using a formal reasoning language or framework).

In our examples we have emphazised single vs. multiple key-sampling 
for randomized countermeasures and discussed its impact on the bounds. 
Such a fine-grained re-randomization for many countermeasures and memory 
layouts has been systematically implemented by 
Curtsinger and Berger~\cite{curtsinger2013stabilizer}.

To the best of our knowledge, we are the first to propose game-based proofs 
for memory-safety countermeasures and to apply the generic approach to several 
countermeasures and their composition.


\section{Conclusions}
\label{sec:conclusions}

In this paper we have presented an approach to reason about probabilistic 
countermeasures against memory-safety vulnerabilities in a rigorous way using 
concepts from cryptographic proofs. We have shown that our modeling is 
applicable for a wide range of countermeasures and their composition. Moreover 
we  have shown how to close certain side-channels using replicas, similarly as 
in Secure Multi-Execution (SME). Surprisingly, 
using this technique also hardens significantly several probabilistic 
countermeasures, by making exploitation 
(theoretically) infeasible.  In future 
work, we plan to model other diversity inspired countermeasures from the 
literature and reason about their guarantees, such as DieHarder \cite{novark2010dieharder}. 
Moreover we plan to use existing tool support to rigorously develop 
computer-aided proofs of our bounds.

\bibliographystyle{IEEEtran}
\bibliography{biblio}

\begin{thebibliography}{10}
\providecommand{\url}[1]{#1}
\csname url@samestyle\endcsname
\providecommand{\newblock}{\relax}
\providecommand{\bibinfo}[2]{#2}
\providecommand{\BIBentrySTDinterwordspacing}{\spaceskip=0pt\relax}
\providecommand{\BIBentryALTinterwordstretchfactor}{4}
\providecommand{\BIBentryALTinterwordspacing}{\spaceskip=\fontdimen2\font plus
\BIBentryALTinterwordstretchfactor\fontdimen3\font minus
  \fontdimen4\font\relax}
\providecommand{\BIBforeignlanguage}[2]{{%
\expandafter\ifx\csname l@#1\endcsname\relax
\typeout{** WARNING: IEEEtran.bst: No hyphenation pattern has been}%
\typeout{** loaded for the language `#1'. Using the pattern for}%
\typeout{** the default language instead.}%
\else
\language=\csname l@#1\endcsname
\fi
#2}}
\providecommand{\BIBdecl}{\relax}
\BIBdecl

\bibitem{younan2012runtime}
Y.~Younan, W.~Joosen, and F.~Piessens, ``Runtime countermeasures for code
  injection attacks against c and c++ programs,'' \emph{ACM Computing Surveys
  (CSUR)}, vol.~44, no.~3, p.~17, 2012.

\bibitem{shacham2007geometry}
H.~Shacham, ``The geometry of innocent flesh on the bone: Return-into-libc
  without function calls (on the x86),'' in \emph{Proceedings of the 14th ACM
  conference on Computer and communications security}.\hskip 1em plus 0.5em
  minus 0.4em\relax ACM, 2007, pp. 552--561.

\bibitem{bittau2014hacking}
A.~Bittau, A.~Belay, A.~Mashtizadeh, D.~Mazieres, and D.~Boneh, ``Hacking
  blind,'' in \emph{Security and Privacy (SP), 2014 IEEE Symposium on}.\hskip
  1em plus 0.5em minus 0.4em\relax IEEE, 2014, pp. 227--242.

\bibitem{snow2013just}
K.~Z. Snow, F.~Monrose, L.~Davi, A.~Dmitrienko, C.~Liebchen, and A.-R. Sadeghi,
  ``Just-in-time code reuse: On the effectiveness of fine-grained address space
  layout randomization,'' in \emph{Security and Privacy (SP), 2013 IEEE
  Symposium on}.\hskip 1em plus 0.5em minus 0.4em\relax IEEE, 2013, pp.
  574--588.

\bibitem{durumeric2014matter}
Z.~Durumeric, J.~Kasten, D.~Adrian, J.~A. Halderman, M.~Bailey, F.~Li,
  N.~Weaver, J.~Amann, J.~Beekman, M.~Payer \emph{et~al.}, ``The matter of
  heartbleed,'' in \emph{Proceedings of the 2014 Conference on Internet
  Measurement Conference}.\hskip 1em plus 0.5em minus 0.4em\relax ACM, 2014,
  pp. 475--488.

\bibitem{abadi2005control}
M.~Abadi, M.~Budiu, U.~Erlingsson, and J.~Ligatti, ``Control-flow integrity,''
  in \emph{Proceedings of the 12th ACM conference on Computer and
  communications security}.\hskip 1em plus 0.5em minus 0.4em\relax ACM, 2005,
  pp. 340--353.

\bibitem{zeng2013strato}
B.~Zeng, G.~Tan, and {\'U}.~Erlingsson, ``Strato: A retargetable framework for
  low-level inlined-reference monitors.'' in \emph{Usenix Security}, 2013, pp.
  369--382.

\bibitem{goktas2014out}
E.~G{\"o}ktas, E.~Athanasopoulos, H.~Bos, and G.~Portokalidis, ``Out of
  control: Overcoming control-flow integrity,'' in \emph{2014 IEEE Symposium on
  Security and Privacy}.\hskip 1em plus 0.5em minus 0.4em\relax IEEE, 2014, pp.
  575--589.

\bibitem{bhatkar2003address}
S.~Bhatkar, D.~C. DuVarney, and R.~Sekar, ``Address obfuscation: An efficient
  approach to combat a broad range of memory error exploits.'' in \emph{USENIX
  Security}, vol.~3, 2003, pp. 105--120.

\bibitem{cowan1999protecting}
C.~Cowan, S.~Beattie, R.~F. Day, C.~Pu, P.~Wagle, and E.~Walthinsen,
  ``Protecting systems from stack smashing attacks with stackguard,'' in
  \emph{Linux Expo}.\hskip 1em plus 0.5em minus 0.4em\relax Citeseer, 1999.

\bibitem{isetrandomization}
G.~S. Kc, A.~D. Keromytis, and V.~Prevelakis, ``Countering code-injection
  attacks with instruction-set randomization,'' in \emph{Proc.\ of the 10th ACM
  conference on Computer and communications security}, 2003.

\bibitem{cowan2003pointguard}
C.~Cowan, S.~Beattie, J.~Johansen, and P.~Wagle, ``Pointguard tm: protecting
  pointers from buffer overflow vulnerabilities,'' in \emph{Proceedings of the
  12th conference on USENIX Security Symposium}, vol.~12, 2003, pp. 91--104.

\bibitem{weiss2006known}
Y.~Weiss and E.~G. Barrantes, ``Known/chosen key attacks against software
  instruction set randomization,'' in \emph{Computer Security Applications
  Conference, 2006. ACSAC'06. 22nd Annual}.\hskip 1em plus 0.5em minus
  0.4em\relax IEEE, 2006, pp. 349--360.

\bibitem{shacham2004effectiveness}
H.~Shacham, M.~Page, B.~Pfaff, E.-J. Goh, N.~Modadugu, and D.~Boneh, ``On the
  effectiveness of address-space randomization,'' in \emph{Proc.\ of the 11th
  ACM conference on Computer and communications security}.\hskip 1em plus 0.5em
  minus 0.4em\relax ACM, 2004, pp. 298--307.

\bibitem{bellare2006security}
M.~Bellare and P.~Rogaway, ``The security of triple encryption and a framework
  for code-based game-playing proofs,'' in \emph{Advances in
  Cryptology-EUROCRYPT 2006}.\hskip 1em plus 0.5em minus 0.4em\relax Springer,
  2006, pp. 409--426.

\bibitem{goldreich1999foundations}
O.~Goldreich, ``The foundations of modern cryptography,'' in \emph{Modern
  Cryptography, Probabilistic Proofs and Pseudorandomness}.\hskip 1em plus
  0.5em minus 0.4em\relax Springer, 1999, pp. 1--37.

\bibitem{DBLP:conf/ecoop/GaneshBO15}
V.~Ganesh, S.~Banescu, and M.~Ochoa, ``Short paper: The meaning of
  attack-resistant systems,'' in \emph{Proceedings of the 10th {ACM} Workshop
  on Programming Languages and Analysis for Security, PLAS@ECOOP 2015, Prague,
  Czech Republic, July 4-10, 2015}, 2015, pp. 49--55.

\bibitem{barthe2009formal}
G.~Barthe, B.~Gr{\'e}goire, and S.~Zanella~B{\'e}guelin, ``Formal certification
  of code-based cryptographic proofs,'' \emph{ACM SIGPLAN Notices}, vol.~44,
  no.~1, pp. 90--101, 2009.

\bibitem{devriese2010noninterference}
D.~Devriese and F.~Piessens, ``Noninterference through secure
  multi-execution,'' in \emph{Security and Privacy (SP), 2010 IEEE Symposium
  on}.\hskip 1em plus 0.5em minus 0.4em\relax IEEE, 2010, pp. 109--124.

\bibitem{diehard}
\BIBentryALTinterwordspacing
E.~D. Berger and B.~G. Zorn, ``Diehard: Probabilistic memory safety for unsafe
  languages,'' in \emph{Proc.\ of the 2006 ACM SIGPLAN Conference on
  Programming Language Design and Implementation}.\hskip 1em plus 0.5em minus
  0.4em\relax ACM, 2006, pp. 158--168. [Online]. Available:
  \url{http://doi.acm.org/10.1145/1133981.1134000}
\BIBentrySTDinterwordspacing

\bibitem{wagle2003stackguard}
P.~Wagle and C.~Cowan, ``Stackguard: Simple stack smash protection for gcc,''
  in \emph{Proceedings of the GCC Developers Summit}.\hskip 1em plus 0.5em
  minus 0.4em\relax Citeseer, 2003, pp. 243--255.

\bibitem{papadogiannakis2013asist}
A.~Papadogiannakis, L.~Loutsis, V.~Papaefstathiou, and S.~Ioannidis, ``Asist:
  architectural support for instruction set randomization,'' in
  \emph{Proceedings of the 2013 ACM SIGSAC conference on Computer \&
  communications security}.\hskip 1em plus 0.5em minus 0.4em\relax ACM, 2013,
  pp. 981--992.

\bibitem{bletsch2011jump}
T.~Bletsch, X.~Jiang, V.~W. Freeh, and Z.~Liang, ``Jump-oriented programming: a
  new class of code-reuse attack,'' in \emph{Proceedings of the 6th ACM
  Symposium on Information, Computer and Communications Security}.\hskip 1em
  plus 0.5em minus 0.4em\relax ACM, 2011, pp. 30--40.

\bibitem{sovarel2005s}
A.~N. Sovarel, D.~Evans, and N.~Paul, ``Where's the feeb? the effectiveness of
  instruction set randomization.'' in \emph{Usenix Security}, 2005.

\bibitem{x86_instr_count}
S.~Heule, ``How many x86-64 instructions are there anyway?''
  \url{https://stefanheule.com/blog/how-many-x86-64-instructions-are-there-anyway/},
  accessed: 2017-01-12.

\bibitem{follner2016analyzing}
A.~Follner, A.~Bartel, and E.~Bodden, ``Analyzing the gadgets towards a metric
  to measure gadget quality,'' \emph{arXiv preprint arXiv:1605.08159}, 2016.

\bibitem{pi32010scraps}
A.~'pi3' Zabrocki, ``Scraps of notes on remote stack overflow exploitation,''
  November 2010.

\bibitem{pucella2010independence}
R.~Pucella and F.~B. Schneider, ``Independence from obfuscation: A semantic
  framework for diversity,'' \emph{Journal of Computer Security}, vol.~18,
  no.~5, pp. 701--749, 2010.

\bibitem{abadi2012protection}
M.~Abadi and G.~D. Plotkin, ``On protection by layout randomization,''
  \emph{ACM Transactions on Information and System Security (TISSEC)}, vol.~15,
  no.~2, p.~8, 2012.

\bibitem{novark2010dieharder}
G.~Novark and E.~D. Berger, ``Dieharder: securing the heap,'' in
  \emph{Proceedings of the 17th ACM conference on Computer and communications
  security}.\hskip 1em plus 0.5em minus 0.4em\relax ACM, 2010, pp. 573--584.

\bibitem{curtsinger2013stabilizer}
C.~Curtsinger and E.~D. Berger, ``Stabilizer: statistically sound performance
  evaluation,'' in \emph{ACM SIGARCH Computer Architecture News}, vol.~41,
  no.~1.\hskip 1em plus 0.5em minus 0.4em\relax ACM, 2013, pp. 219--228.

\end{thebibliography}

\end{document}